# Revisiting Trends in the Exchange Current for Hydrogen Evolution

Timothy T. Yang[1], Rituja B. Patil[2], James R. McKone[2], and Wissam A. Saidi[1*]

[1]Department of Mechanical Engineering and Materials Science, University of Pittsburgh, Pittsburgh, PA 15260, United States

[2]Department of Chemical and Petroleum Engineering, University of Pittsburgh, Pittsburgh, PA 15260, United States

*To whom correspondence should be addressed: alsaidi@pitt.edu

## Abstract

Nørskov and collaborators proposed a simple kinetic model to explain the volcano relation for the hydrogen evolution reaction on transition metal surfaces in such that $j_0 = k_0 f(\Delta G_H)$ where $j_0$ is the exchange current density, $f(\Delta G_H)$ is a function of the hydrogen adsorption free energy $\Delta G_H$ as computed from density functional theory, and $k_0$ is a universal rate constant. Herein, focusing on the hydrogen evolution reaction in acidic medium, we revisit the original experimental data and find that the fidelity of this kinetic model can be significantly improved by invoking metal-dependence on $k_0$ such that the logarithm of $k_0$ linearly depends on the absolute value of $\Delta G_H$. We further confirm this relationship using additional experimental data points obtained from a critical review of the available literature. Our analyses show that the new model decreases the discrepancy between calculated and experimental exchange current density values by up to four orders of magnitude. Furthermore, we show the model can be further improved using machine learning and statistical inference methods that integrate additional material properties.

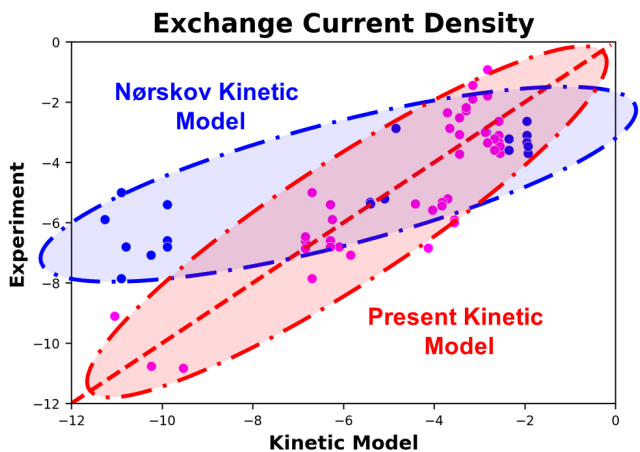

Hydrogen is a powerful energy carrier that can be generated through the hydrogen evolution reaction (HER) – the critical cathodic reaction of electrochemical water-splitting. Not surprisingly, HER is still a topic of great fundamental interest for electrochemical energy conversion.[1-9] To understand the HER activity, Bockris[10], Conway and Bockris[11], Petrenko[12], Kita[13] and Trasatti[14-16] correlated experimental HER reaction rates with physicochemical descriptors such as atomic number, work function, d-band center, the heat of hydrogen adsorption, and Pauling electronegativities. Motivated by Sabatier's principle that the maximum catalytic rate is achieved when the interaction between the reactants and catalyst is neither too strong nor too weak,[17] Parsons and Gerischer independently proposed the free energy of hydrogen adsorption, $\Delta g^0$, as an HER descriptor such that the maximum rate corresponds to a minimum in the magnitude of $\Delta g^0$ under equilibrium conditions.[18,19] To link theoretical $\Delta g^0$ with experimental measurements, early attempts were made by Krishtalik and Trasatti.[15] Later, Krishtalik[20] and Trasatti[15] used a compilation of experimental data to validate Parsons volcano relationship using the metal-hydrogen interaction strength based on Eley-Stevenson method.[21, 22] However, this approach had limited success because the maximum exchange current was not associated with $\Delta g^0 = 0$, as proposed by Parsons.

In a seminal study, Nørskov and collaborators introduced the hydrogen adsorption free energy $\Delta G_H$ computed from density functional theory (DFT) as an accurate estimate of Parsons' $\Delta g^0$. Further, this study built a connection with electrochemical exchange currents using a simple kinetic model based on systematic investigations on transition metal surfaces.[23] We will refer to this model as the Nørskov model hereafter. The Nørskov model confirms the theoretical volcano trend proposed by Parsons where $\Delta G_H \approx 0$ at the maximum exchange current density of the HER. Further, this model also demonstrated that a computational framework based on an easy-to-compute descriptor $\Delta G_H$ can provide a rational approach to catalyst design, which improved on approaches to materials design based on trial-and-error or chemical heuristics that have historically been the norm for experimental catalysis research.

Although the Nørskov model for the HER has been widely accepted by the electrochemistry community (e.g., see recent studies[24-27]), several studies highlighted caveats in this model. For example, the Nørskov model is applied to pristine metallic surfaces to compute $\Delta G_H$ while under electrochemical conditions, many metals are likely to be oxidized or exhibit amorphous surface structure.[28-30] Further, it was argued that electrostatic effects from metal-water interfaces and the effects from the adsorption of water and oxygen, as well as kinetic factors and d-band characteristics, are not adequately accounted for in the Nørskov model.[28-30] In addition, it has been recently argued that Pt is not a thermoneutral catalyst with $\Delta G_H$ that deviates from zero.[31-33] Several studies proposed HER models based on microkinetic analysis.[34, 35] However, these models are generally complex with many parameters that are obtained by some approximation for by fitting to experimental data. *These caveats notwithstanding, the Nørskov model remains a leading framework for the design of HER catalysts and heterogeneous electrocatalysts in general.*[1-5]

Herein we revisit the Nørskov model and show that the calculated exchange current densities deviate from the corresponding experimental values by up to six orders of magnitude.[23] While differences between experimental and DFT-computed rates are not uncommon, we show that the discrepancy can be substantially reduced by considering that the rate constant is material-dependent rather than universal, as assumed in the original Nørskov model[23]. Specifically, we present evidence that the kinetic pre-factor $k_0$ also depends on the absolute value of $\Delta G_H$. We further validate the findings based on reliable data obtained from a critical review of experimental exchange current densities on transition metal surfaces from the available research literature.

In the Nørskov model, the exchange current density $j_0$, which describes the magnitude of the forward and reverse reaction rates at equilibrium, is defined as

$$j_0 = \begin{cases} ek_0 \exp(-\Delta G_H/k_B T)C_{tot}(1-\theta) & \text{for } \Delta G_H > 0 \\ ek_0 C_{tot}(1-\theta) & \text{for } \Delta G_H < 0 \end{cases}, \qquad (1)$$

where $e$ is the charge of an electron and $C_{tot}$ is the areal concentration of adsorption sites. Note that while the Nørskov model was presented in terms of the exchange current (e.g., amps), Eq. (1) is defined in terms of the exchange current density (e.g., amps per square cm), which necessitates the inclusion of an area-normalized $C_{tot}$. The model is derived from the basic relation that the current is linearly proportional to the concentration of reactants $H^+$ ($H^*$) for $\Delta G_H > 0$ ($\Delta G_H < 0$). Further, $k_0$ encompasses several factors such as additional concentration factors due to applied and formal potentials, reaction rates for all elementary steps, and the effects from transfer coefficients in microkinetic models.[34-36] Based on ab initio thermodynamic analysis,[37] we assume that the hydrogen coverage is at the lowest limit for metal surfaces that repel hydrogen (i.e., with $\Delta G_H > 0$), and attains a full monolayer coverage for surfaces that are attractive to hydrogen (i.e., with $\Delta G_H < 0$). At a given temperature T, the fraction of surface occupied sites by hydrogen follows the Langmuir model such that $\theta = K/(1+K)$ with $K = \exp(-\Delta G_H/k_B T)$. $k_0$ is the rate constant that is assumed to have a universal value for all metals, which is taken as $k_0 = 200\ s^{-1}\text{site}^{-1}$ by linearly fitting exchange currents to experimental data.[23] In a previous study we have shown that $k_0$ assumes a different value for β-Mo₂C nanoparticles by fitting to experimental results.[38]

The hydrogen adsorption free energy $\Delta G_H$ is obtained from the free energy difference between the hydrogen in gas phase and in adsorbed phase, which can be computed as,

$$\Delta G_H = \Delta E_{ads} + \Delta E_{ZPE} - T\Delta S, \qquad (2)$$

where $\Delta E_{ZPE}$ is the zero-point energy that is found to be less than 0.05 eV for all metals, consistent with prior results.[23] $\Delta S$ is the entropy between the hydrogen adsorbed state and the gas state. It can be approximated as $-\frac{1}{2}S^0_{H_2}$ where $S^0_{H_2} = 1.35 \times 10^{-3}$ eV/K is the entropy of hydrogen in the gas phase at room temperature, as obtained from experimental measurements.[39] The hydrogen adsorption energy $\Delta E_{ads}$ is defined as,

$$\Delta E_{ads} = \frac{1}{n}\left(E_{slab/nH^*} - E_{slab} - \frac{n}{2}E_{H_2}\right), \qquad (3)$$

where $n$ is the number of adsorbed hydrogen atoms, $\mathrm{E_{slab/nH^*}}$ and $\mathrm{E_{slab}}$ are the energies of the slab with $n$ adsorbed hydrogen atoms ($\mathrm{H^*}$) and of a clean slab respectively, and $\mathrm{E_{H_2}}$ is the energy of $\mathrm{H_2}$ in gas phase. All terms in Eq. (3) are directly computed from DFT. The Nørskov model of Eq. (1) shows that the maximum catalytic activity is at $\Delta \mathrm{G_H} = 0$ and the activity decreases when $\Delta \mathrm{G_H}$ moves away from zero, thus reproducing the volcano relationship for exchange current and adsorption free energy $\Delta \mathrm{G_H}$.

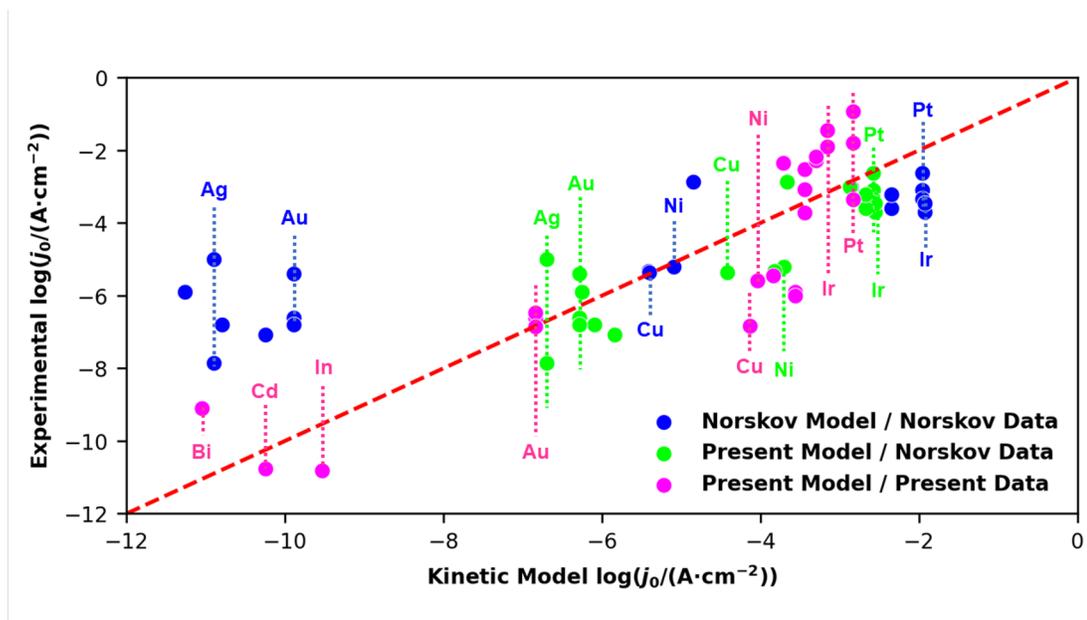

**Figure 1.** Comparison of experimental and calculated exchange current densities. We use the experimental data provided from the original work of Ref. 14 (Nørskov Data) and our database (Present Data). The models from the original work (Nørskov Model) and our modified model (Present Model) are used for the calculation of $j_0$. To avoid clutter, we only labeled few elements belonging to three categories that are excellent (Pt and Ir), moderate (Ni and Cu) and poor (Ag, Au, Bi, In and Cd) HER catalysts. All data are shown in Table S3.

The data labeled "Nørskov Model / Nørskov Data" in Figure 1 shows the experimental exchange current densities $j_0$ obtained from the original report vs. the computed values based on the Nørskov model. For consistency, we use the same $\Delta \mathrm{G_H}$ values as in the original report to compute $j_0$ from Eq. (1). As seen in Figure 1, the Nørskov model qualitatively captures the experimental trend for the currents among the different metals. However, the model underestimates the experimental exchange currents by 3 – 6 orders of magnitude for the metal surfaces with low HER activity such as W, Nb, Au, and Ag, while it is in better agreement (within two orders of magnitude) for the highly catalytic surfaces of Pt, Pd, Ir, and Rh. We further confirm that the underestimation of exchange currents on low activity surfaces is not due to experimental errors. For example, for Bi, In and Cd, the experimental values are $\sim 10^{-10}$ A/cm$^2$ that we have carefully examined from available literature (see Table S5). In contrast, the corresponding exchange currents obtained from Nørksov model are $\sim 10^{-21}$, $\sim 10^{-17}$ and $\sim 10^{-19}$ A/cm$^2$, respectively. The large discrepancy between experimental and theoretical predictions, in addition to the systematic variation in the discrepancy, strongly suggests that the model does not capture one or more relevant physical parameters.

In the Nørskov model, the rate constant $k_0$ is assumed to be constant for all metals, which was justified considering that $k_0$ includes mainly effects of solvent reorganization during proton transfer to electrode surfaces. However, this approximation is too simplistic given the significant differences in the HER kinetics between highly efficient surfaces such as Pt and Ir, and surfaces with lower HER activity such as W and Ag. For example, a very weak H-binding surface is likely to have a transition state that is similar in nature to a surface-bound hydrogen (the Volmer reaction as the rate-determining step), whereas a strong H-binding surface is likely to have a transition state that resembles H$^+$ in the electrolyte or a weakly bound H$_2$ (the Heyrovsky or the Tafel reaction is rate-determining). Thus, outer-sphere effects that can be rationalized as the reorganization of solvents[40] may be more significant for strong H-binding surfaces than weak H-binding surfaces.[41] Further, other factors contribute to different HER dynamics between metal surfaces, such as the solution pH and the composition and structure of the double layer, particularly for inefficient catalysts that require very large overpotentials (and commensurately large electric fields across the double layer) to drive the HER.[42]

Re-examining the data provided in the original report[23], our analyses clearly show that a universal $k_0$ value is not justified. Figure 2 presents implied $k_0$ values obtained from linear regression analysis of Eq. (1) using experimental $j_0$ values and the DFT computed $\Delta G_\text{H}$. This approach amounts to testing the hypothesis that deviations between the experimental and DFT-predicted $j_0$ using the Nørskov Model can be attributed to a metal-dependent value of $k_0$. As seen in the figure, for both low and high coverage limits, $k_0$ varies systematically with $\Delta G_H$ over several orders of magnitude, where metals that weakly interact with hydrogen have low $k_0$ values and those that strongly interact with hydrogen, either repulsive or attractive, have larger $k_0$ values. Further, as shown in Figure 2, we find a strong linear correlation between $\ln(k_0)$ and $|\Delta G_H|$. Indeed, invoking this linear relationship is found to substantially decrease the inconsistency between the kinetic model and the experimental values, as shown in Figure 1 for the data labeled "Present Model / Nørskov Data". We conclude from the analysis based on data in Ref. [23] that $k_0$ is not universal but is material-specific, at least to the extent that efficient versus inefficient metals should exhibit characteristically different $k_0$ values. Notably, when restricted to a model that only incorporates metals with a similar range of H-binding energies, this effect is diminished – hence the fit is relatively good near the peak of the volcano only.

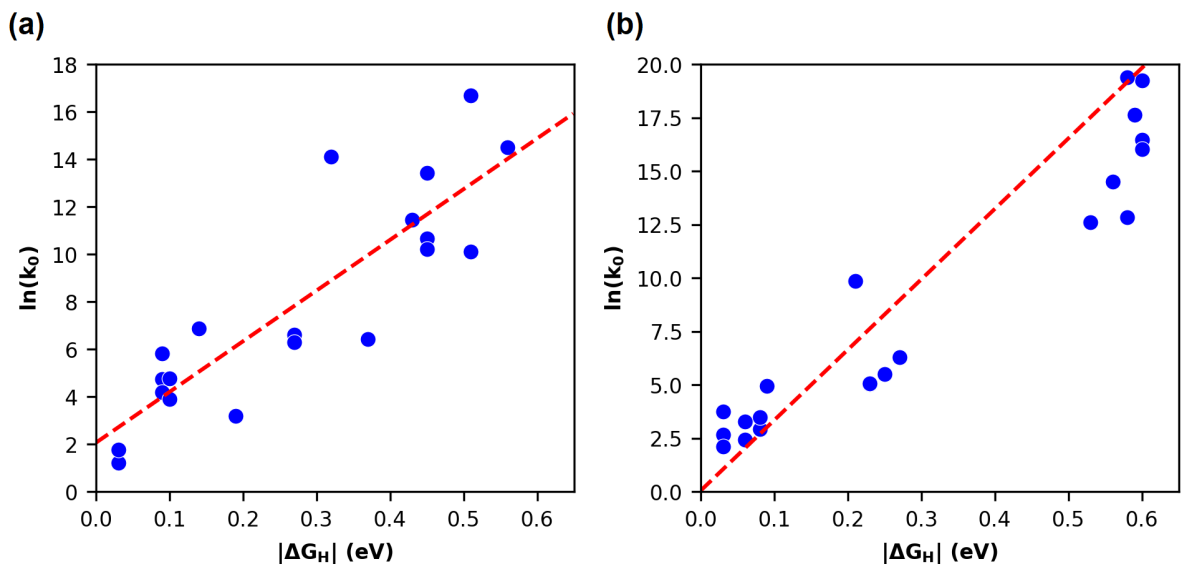

**Figure 2.** The correlation of $\ln(k_0)$ vs. absolute value for the hydrogen adsorption free energy $|\Delta G_\text{H}|$ calculated from DFT for (a) low- and (b) high-coverage limits. All data including experimental currents and computed $\Delta G_\text{H}$ are from Ref. 14.

Recently, the Brønsted–Evans–Polanyi (BEP) relation, a linear relation between a reaction's free energy and its activation energy $E_a$, is confirmed for the HER on pure metal surfaces[43] using a computational approach that corrects for finie-size effects in periodic supercell simulations. For example, for the metals with $\Delta G_H < 0$ ($\Delta G_H > 0$), the activation barrier of the rate-limiting Heyrovsky (Volmer) reaction decreases with increasing (decreasing) $\Delta G_H$. The BEP relation on HER is also confirmed experimentally on precious metals of Pt, Ir, Pd and Rh.[44] If we assume that $k_0$ of the Nørskov model is the HER reaction rate constant, it follows from Arrhenius relation that $\ln(k_0)$ is linearly proportional to activation energy $E_a$, and in conjunction with the BEP relationship, we can infer the linear dependence between $\ln(k_0)$ and $|\Delta G_H|$ that is obtained using our data-driven approach. Further, the enthalpy-entropy compensation where $E_a$ (or $\Delta G_H$) has a linear relationship with entropy[45] suggests that $\ln(k_0)$ includes effects from activation entropy that is metal-dependent[46], which also supports our findings. However, we believe that a careful derivation is needed to formally derive the dependence between $k_0$ and $E_a$.

To further examine the metal-dependence of $k_0$, we have compiled an additional set of experimental $j_0$ values from a thorough literature search (see comments and references in Table S5) and re-analyzed $\Delta G_H$ using different DFT functionals. The comparison between experimental and present model data are shown in Figure 1 under "Present Model / Present Data". Figure 3(a) shows the correlation between $\ln(k_0)$ and $|\Delta G_H|$ and Figure 3(b) shows the volcano relationship corresponding to the new data. The DFT calculations are carried out using the Vienna Ab Initio Simulation Package (VASP).[47-49] More details about the DFT computational framework are provided in the Supporting Info. We employ the conventional[50] (PBE) and revised[51] (RPBE) Perdew-Burke-Ernzerhof exchange-correlational functional with and without van der Waals (vdW) corrections[52, 53] to assess the variability of the results with the computational framework. The results shown in Figure 1 and Figure 3 are based on RPBE+vdW; however, the findings are found not to be sensitive to the functional.

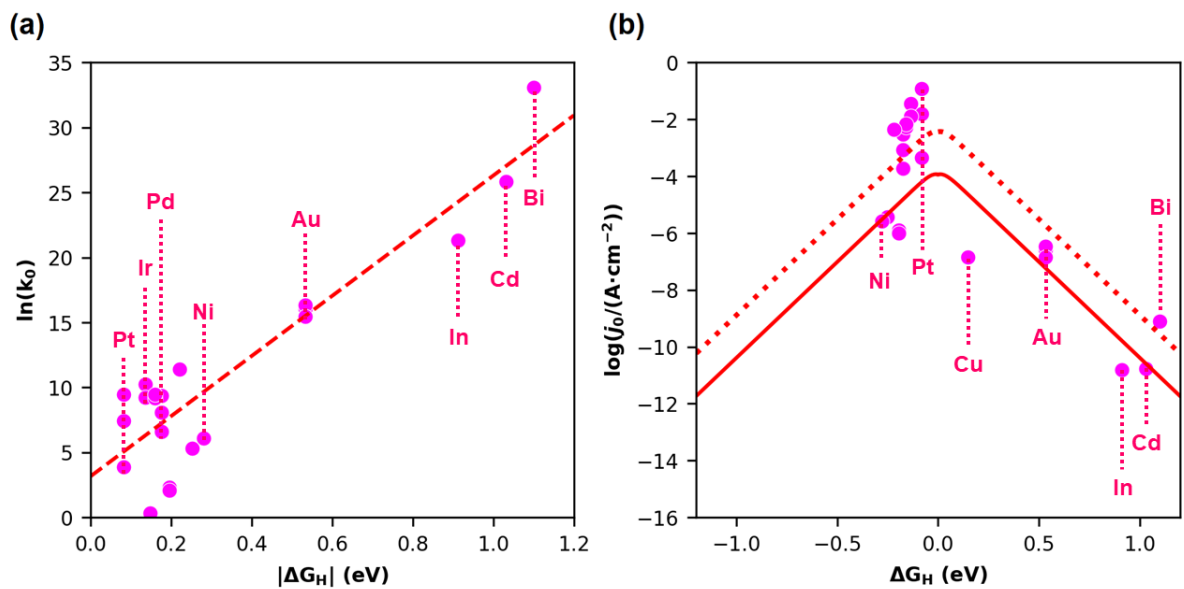

**Figure 3.** (a) Correlation between $\ln(k_0)$ and $|\Delta G_H|$ using "Present Data" of Figure 1 (see Table S4). From the linear fitting, we obtain $\ln(k_0) = 23.16|\Delta G_H| + 3.17$ with a correlation factor of $r^2 = 0.82$. (b) The new volcano curve of $j_0$ based on the Present Model / Present Data. The dotted and solid red lines are obtained using Eq. (1) with the maximum and the minimum $C_{tot}$ in our data, respectively.

The library of experimental exchange current densities[54-68] labeled as "Present Data" in Figure 1 and used in Figure 3 was collected from a larger set of prior literature reports, which were then down-selected to only those reports that minimized or otherwise accounted for the impacts of electrolyte/surface contamination, electrode roughness, and mass transfer effects. Table 1 summarizes experimental results. A detailed discussion of this down-selection process is included in the Supporting Info. We also chose to exclude metals that are expected to be oxidized under HER conditions in acidic solution, such as Mo, W, and Nb.[69] Finally, we added measurements on Cd, In, and Bi and Ru that were not included in Ref. [23]. Figure 3 shows that considerably more experimental data are available for HER-active metals, which accounts for the greater density of points with small $|\Delta G_H|$. The spread in these data likely reflects uncertainty both in DFT-calculated $\Delta G_H$ and experimental $j_0$.[70] Nonetheless, as shown in Figure 3 (a), there is a clear linear relationship between $\ln k_0$ and $|\Delta G_H|$ that can be described as $\ln(k_0) = 23.16|\Delta G_H| + 3.17$. This is specifically evidenced in clustering of three metal groups with characteristically different H binding energies: the precious metals (Pt, Ir, Pd and Rh) near $|\Delta G_H| = 0.1$ eV, the metals near $|\Delta G_H| = 0.5$ eV, and the HER inert metals near $|\Delta G_H| = 1$ eV.

| Table 1. The collected exchange current densities from experiments. | | | | |
|---|---|---|---|---|
| **Electrode** | **Reported $j_0$ (A/cm$^2$)** | **Electrolyte** | **Temperature** | **Reference** |
| Pt (111) | 4.5×10$^{-4}$ | 0.05 M H$_2$SO$_4$ | 303 K | 54 |
| Pt (100) | 6.0×10$^{-4}$ | | | |
| Pt (110) | 9.8×10$^{-4}$ | | | |
| Pt/C | 1.6×10$^{-2}$ | 0.2 M H$_3$PO$_4$ | 293 K | 55 |
| Pt/C | 1.2×10$^{-1}$ | 0.1 M HClO$_4$ | 313 K | 56 |
| Ir/C | 3.6×10$^{-2}$ | 0.1 M HClO$_4$ | 313 K | 56 |
| Ir/C | 1.28×10$^{-2}$ | 0.2 M H$_2$SO$_4$ | 293 K | 55 |
| Pd | 1.9×10$^{-4}$ | 0.5 M H$_2$SO$_4$ | Not mentioned | 57 |
| Pd/C | 3.0×10$^{-3}$ | 0.1 M HClO$_4$ | 313 K | 56 |
| Pd/C | 8.4×10$^{-4}$ | 0.1 M HClO$_4$ | 293 K | 55 |
| Rh/C | 5.2×10$^{-3}$ | 0.1 M HClO$_4$ | 313 K | 56 |
| Rh/C | 6.7×10$^{-3}$ | 0.1 M HClO$_4$ | 293 K | 55 |
| Ru | 4.5×10$^{-3}$ | 1M HCl + NaCl | 298 K | 58 |
| Cu | 1.45×10$^{-7}$ | 0.1 N HCl | Not mentioned | 59 |
| Co | 3.6×10$^{-6}$ | 1 M H$_2$SO$_4$ | 293 K | 60 |
| Ni | 2.6×10$^{-6}$ | 0.5 M H$_2$SO$_4$ | 295 K | 61 |
| Au (111) | 2.5×10$^{-7}$ | 0.1 M HClO$_4$ | Not mentioned | 62 |
| Au (100) | 0.5×10$^{-7}$ | | | |
| Au (110) | 0.3×10$^{-7}$ | | | |
| Au (111) | 3.38×10$^{-7}$ | 0.5 M H$_2$SO$_4$ | Not mentioned | 63 |
| Poly Au | 1.40×10$^{-7}$ | 0.1 M HClO$_4$ | Not mentioned | 62 |
| Re | 1.25×10$^{-6}$ | 0.5 M H$_2$SO$_4$ | 298 K | 64 |
| Re | 1×10$^{-6}$ | 0.5 M H$_2$SO$_4$ | 298 K | 65 |
| Cd | 1.7×10$^{-11}$ | 0.5 N H$_2$SO$_4$ | Not mentioned | 66 |
| Bi | 8×10$^{-10}$ | 4.8 M H$_2$SO$_4$ | Not mentioned | 67 |
| In | 1.51×10$^{-11}$ | 0.1 M HClO$_4$ | 303 K | 68 |

Our results clearly show that $\Delta G_H$ more accurately describes $j_0$ for the HER if we include an additional exponential relationship between $|\Delta G_H|$ on $k_0$. However, it is conceivable that $k_0$, and hence the exchange current density, also depends on other metal properties besides $\Delta G_H$. For instance, previous studies postulated that the HER rate can be modeled by using atomic number, work functions, and Pauling electronegativities as material descriptors.[10-16] To investigate this, we used a machine learning approach based on SISSO (sure independence screening and sparsifying operator)[71-74] to develop an accurate and physically interpretable model for $\ln k_0$. We investigated the following primary atomic features in this analysis: empirical radius, mass, number, period in Periodic Table, electron affinity, ionization energy, and Pauling electronegativity, in addition to the following metal features: density and work function. In the SISSO approach, potential descriptors for $\ln k_0$ are formed from the primary features with up to ten level interactions of complexity utilizing three mathematical operations (addition, multiplication, and division). The limited size of the experimental dataset (12 metals) precludes a full investigation, and thus, we used a relatively small number of primary features and mathematical operations in the construction of only one-dimensional descriptors. By searching the massive space of potential descriptors, SISSO identified many models for $\ln k_0$ that capture a large proportion of the variations among different elements. Table S6 lists the best ten models with correlation coefficients $r^2 > 0.975$. Notably all these models are found to depend on $|\Delta G_H|$ indicating its prime effect on $k_0$. However, a larger experimental database is needed to unambiguously validate the findings, and to identify other, if any, important material properties that affect $k_0$.

In summary, we agree with the original work by Nørskov and collaborators that the trend of $j_0$ can be explained by a kinetic model that relies on $\Delta G_H$ as the sole descriptor. However, after carefully analyzing the experimental and computational results, we propose that the same kinetic model better matches with experiments over a wide range of metals by treating the logarithm of the rate constant $k_0$ as a linear function of the absolute value of $\Delta G_H$.

**Acknowledgements**

W. A. S. and T. Y. acknowledge partial financial support from the National Science Foundation (Award No. DMR-1809085). R.B.P. and J.R.M. acknowledge the Arnold and Mabel Beckman Foundation for financial support via the Beckman Young Investigator program. We are grateful for computing time provided in part by the CRC resources at the University of Pittsburgh and Argonne Leadership Computing Facility, which is a DOE Office of Science User Facility supported under Contract DE-AC02-06CH11357.**References**

# Supporting Information:
# Revisiting Trends in the Exchange Current for Hydrogen Evolution

*Timothy T. Yang[1], Rituja B. Patil[2], James R. McKone[2], and Wissam A. Saidi[1]*

[1]Department of Mechanical Engineering and Materials Science, University of Pittsburgh, Pittsburgh, PA 15260, United States

[2]Department of Chemical and Petroleum Engineering, University of Pittsburgh, Pittsburgh, PA 15260, United States

*To whom correspondence should be addressed: alsaidi@pitt.edu

## Computational Details

We use the Vienna Ab Initio Simulation Package (VASP) for the first-principles density functional theory (DFT) calculations. We use the Perdew-Burke-Ernzerhof (PBE) exchange-correlational functional and Revised Perdew-Burke-Ernzerhof (RPBE) to solve the Kohn-Sham equations within periodic boundary conditions, and the PAW pseudopotentials to describe electron-nucleus interactions.[1-4] The dDsC dispersion correction is used for the Van der Waals (VdW) corrections. The electronic self-consistent loops are terminated within energy-change tolerance of $1\times10^{-6}$ eV. The periodic slab models are 3×3 supercells cleaved using the calculated bulk structures in Table 1 with a thickness of at least 10 Å. We used 15 Å vacuum perpendicular to the surfaces. The relaxations are done by fixing at least two layers in bulk position and relaxing the top three layers for all surfaces. We use the most stable terminations: for the metals in $Fm\bar{3}m$ and $P6_3/mmc$ phase[5-6], we use (111) and (001), respectively, (111) for Bi and In surfaces[7], and (110) for Mo surface.[8]

    The calculated free energy ($\Delta G_H$) values in Table 2 corresponds to the hydrogen coverages of 1/9 ML or 1 ML for the surfaces with positive or negative $\Delta G_H$, respectively. A single water layer is added to test the solvation effects. We find that the effects on adsorption energies are less than 0.05 eV for all the metals therefore not included. Comparing all the $\Delta G_H$ values for a metal, it is observed that the PBE results are systematically ~0.15 eV less than the results obtained from RPBE for the cases with and without VdW corrections. For each functional, the VdW corrections amounts to less than 0.06 eV. In summary, the trends we show in Figure 3 will be similar for the four cases of $\Delta G_H$ calculations.

| | K-points | Phase | PBE | RPBE |
|---|---|---|---|---|
| **Table S1.** Bulk structures obtained with PBE and RPBE functionals. | | | | |
| Ag | 12×12×12 | $Fm\bar{3}m$ | a=b=c=2.93<br>α=β=γ=60 | a=b=c=2.97<br>α=β=γ=60 |
| Au | 12×12×12 | $Fm\bar{3}m$ | a=b=c=2.94<br>α=β=γ=60 | a=b=c=2.97<br>α=β=γ=60 |
| Bi | 12×12×12 | R$\bar{3}$m | a=b=c=4.83<br>α=β=γ=56.69 | a=b=c=4.99<br>α=β=γ=54.98 |
| Cd | 12×12×12 | P6$_3$/mmc | a=b=3.05; c=5.59<br>α=β=90; γ=120 | a=b=3.10; c=5.67<br>α=β=90; γ=120 |
| Co | 12×12×12 | Fm$\bar{3}$m | a=b=c=2.44<br>α=β=γ=60 | a=b=c=2.46<br>α=β=γ=60 |
| Co | 12×12×12 | P6$_3$/mmc | a=b=2.45; c=3.95<br>α=β=90; γ=120 | a=b=2.47; c=3.98<br>α=β=90; γ=120 |
| Cu | 12×12×12 | Fm$\bar{3}$m | a=b=c=2.57<br>α=β=γ=60 | a=b=c=2.60<br>α=β=γ=60 |
| In | 12×12×12 | R$\bar{3}$m | a=b=c=8.49<br>α=β=γ=23.139 | a=b=c=8.61<br>α=β=γ=23.134 |
| Ir | 12×12×12 | Fm$\bar{3}$m | a=b=c=2.74<br>α=β=γ=60 | a=b=c=2.75<br>α=β=γ=60 |
| Mo | 8×8×8 | Im$\bar{3}$m | a=b=c=2.73<br>α=β=γ=109.471 | a=b=c=2.74<br>α=β=γ=109.471 |
| Ni | 12×12×12 | Fm$\bar{3}$m | a=b=c=2.48<br>α=β=γ=60 | a=b=c=2.50<br>α=β=γ=60 |
| Pd | 12×12×12 | Fm$\bar{3}$m | a=b=c=2.78<br>α=β=γ=60 | a=b=c=2.81<br>α=β=γ=60 |
| Pt | 12×12×12 | Fm$\bar{3}$m | a=b=c=2.81<br>α=β=γ=60 | a=b=c=2.82<br>α=β=γ=60 |
| Re | 12×12×12 | P6$_3$/mmc | a=b=2.77 c=4.48<br>α=β=90; γ=120 | a=b=2.78; c=4.50<br>α=β=90; γ=120 |
| Rh | 12×12×12 | Fm$\bar{3}$m | a=b=c=2.70<br>α=β=γ=60 | a=b=c=2.72<br>α=β=γ=60 |
| Ru | 12×12×12 | P6$_3$/mmc | a=b=2.72; c=4.28<br>α=β=90; γ=120 | a=b=2.73; c=4.30<br>α=β=90.000; γ=120 |
| Nb | 12×12×12 | Im$\bar{3}$m | a=b=c=2.88<br>α=β=γ=109.471 | a=b=c=2.89<br>α=β=γ=109.471 |
| W | 12×12×12 | Im$\bar{3}$m | a=b=c=2.75<br>α=β=γ=109.471 | a=b=c=2.76<br>α=β=γ=109.471 |

**Table S2.** The calculated $\Delta G_H$ from two functionals with and without VdW correction.

| | PBE | RPBE | PBE with VdW | RPBE with VdW |
|---|---|---|---|---|
| Ag | 0.416 | 0.565 | 0.381 | 0.527 |
| Au | 0.418 | 0.574 | 0.388 | 0.532 |
| Bi | 1.040 | 1.127 | 1.004 | 1.099 |
| Cd | 1.051 | 1.038 | 1.058 | 1.031 |
| Co (Fm$\bar{3}$m) | -0.355 | -0.202 | -0.411 | -0.252 |
| Co (P6$_3$/mmc) | -0.352 | -0.201 | -0.407 | -0.250 |
| Cu | 0.040 | 0.188 | -0.002 | 0.147 |
| In | 0.847 | 0.948 | 0.813 | 0.912 |
| Ir | -0.208 | -0.081 | -0.273 | -0.136 |
| Mo | -0.453 | -0.321 | 3.620 | -0.352 |
| Ni | -0.388 | -0.231 | -0.445 | -0.280 |
| Pd | -0.286 | -0.123 | -0.359 | -0.176 |
| Pt | -0.192 | -0.038 | -0.232 | -0.082 |
| Re | -0.293 | -0.154 | -0.333 | -0.195 |
| Rh | -0.270 | -0.116 | -0.318 | -0.159 |
| Ru | -0.330 | -0.180 | -0.375 | -0.220 |

**Table S3-1.** The experimental $\log(j_0/(A \cdot cm^2))$ from Ref [9] (Nørskov Data) is compared with the calculated $\log(j_0/(A \cdot cm^2))$ using Eq. (1) with constant $k_0 = 200$ s$^{-1}$ (Nørskov Model) and with $k_0 = \exp(23.16|\Delta G_H/eV| + 3.17)$ (Present Model).

| | Nørskov Data | Nørskov Model | Present Model |
|---|---|---|---|
| Pt | -3.1 | -1.95 | -2.58 |
| Pt | -2.63 | -1.95 | -2.58 |
| Pt | -3.34 | -1.95 | -2.58 |
| Ir | -3.7 | -1.93 | -2.55 |
| Ir | -3.46 | -1.93 | -2.55 |
| Pd | -3 | -2.85 | -2.87 |
| Pd | -3 | -2.85 | -2.87 |
| Rh | -3.6 | -2.35 | -2.67 |
| Rh | -3.22 | -2.35 | -2.67 |
| Ni | -5.21 | -5.09 | -3.70 |
| Ni | -5.2 | -5.09 | -3.70 |
| Co | -5.32 | -5.41 | -3.82 |
| W | -5.9 | -11.26 | -6.25 |
| W | -5.9 | -11.26 | -6.25 |
| Cu | -5.37 | -5.41 | -4.42 |
| Mo | -7.07 | -10.25 | -5.84 |
| Re | -2.87 | -4.85 | -3.66 |
| Nb | -6.8 | -10.80 | -6.09 |
| Au | -6.6 | -9.89 | -6.29 |
| Au | -6.8 | -9.89 | -6.29 |
| Au | -5.4 | -9.89 | -6.29 |
| Ag | -5 | -10.90 | -6.69 |
| Ag | -7.85 | -10.90 | -6.69 |

**Table S3-2.** The experimental $\log(j_0/(A \cdot cm^2))$ from Table 5 (Present Data) is compared with the calculated $\log(j_0/(A \cdot cm^2))$ with $k_0 = \exp(23.16|\Delta G_H/eV| + 3.17)$ (Present Model). The first column label reflects experimental setup. The "Present Model" utilize slabs of well-defined terminations as described before.

| | Present Data | Present Model |
|---|---|---|
| Pt (111) | -3.35 | -2.83 |
| Pt/C | -1.80 | -2.83 |
| Pt/C | -0.92 | -2.83 |
| Ir | -1.44 | -3.15 |
| Ir | -1.89 | -3.15 |
| Pd | -3.72 | -3.44 |
| Pd | -2.52 | -3.44 |
| Pd | -3.08 | -3.44 |
| Rh/C | -2.28 | -3.30 |
| Rh/C | -2.17 | -3.30 |
| Ru | -2.35 | -3.71 |
| Cu | -6.84 | -4.13 |
| Co | -5.44 | -3.84 |
| Ni | -5.59 | -4.04 |
| Au (111) | -6.60 | -6.84 |
| Au (111) | -6.47 | -6.84 |
| Poly Au | -6.85 | -6.84 |
| Re | -5.90 | -3.56 |
| Re | -6.00 | -3.56 |
| Bi | -9.10 | -11.05 |
| Cd | -10.77 | -10.24 |
| In | -10.82 | -9.54 |

**Table S4.** The data used in Figure 3 in the main paper. The experimental $j_0$s are collected from reliable literatures listed in Table S4. The calculated $\ln(k_0)$ and $\Delta G_H$ are obtained using Eq. (1) and Eq. (2), respectively, in the main document. The area is determined from the structures obtained using RPBE functional listed in Table S1.

| | $\Delta G_H$ | Exp. $j_0$ | log $j_0$ | $\ln(k_0)$ | Area (cm$^2$) | # sites in Area | Ref. |
|---|---|---|---|---|---|---|---|
| Pt (111) | -0.082 | 4.50 × 10$^{-4}$ | -3.35 | 3.87 | 6.20 × 10$^{-15}$ | 9 | 10 |
| Pt/C | -0.082 | 1.60 × 10$^{-2}$ | -1.80 | 7.43 | 6.20 × 10$^{-15}$ | 9 | 11 |
| Pt/C | -0.082 | 1.20 × 10$^{-1}$ | -0.92 | 9.46 | 6.20 × 10$^{-15}$ | 9 | 12 |
| Ir | -0.136 | 3.60 × 10$^{-2}$ | -1.44 | 10.27 | 5.89 × 10$^{-15}$ | 9 | 12 |
| Ir | -0.136 | 1.28 × 10$^{-2}$ | -1.89 | 9.23 | 5.89 × 10$^{-15}$ | 9 | 11 |
| Pd | -0.176 | 1.90 × 10$^{-4}$ | -3.72 | 6.60 | 6.16 × 10$^{-15}$ | 9 | 13 |
| Pd | -0.176 | 3.00 × 10$^{-3}$ | -2.52 | 9.37 | 6.16 × 10$^{-15}$ | 9 | 14 |
| Pd | -0.176 | 8.4 × 10$^{-4}$ | -3.08 | 8.08 | 6.16 × 10$^{-15}$ | 9 | 11 |
| Rh/C | -0.159 | 5.20 × 10$^{-3}$ | -2.28 | 9.20 | 5.77 × 10$^{-15}$ | 9 | 12 |
| Rh/C | -0.159 | 6.70 × 10$^{-3}$ | -2.17 | 9.45 | 5.77 × 10$^{-15}$ | 9 | 11 |
| Ru | -0.220 | 4.50 × 10$^{-3}$ | -2.35 | 11.40 | 5.81 × 10$^{-15}$ | 9 | 15 |
| Cu | 0.147 | 1.45 × 10$^{-7}$ | -6.84 | 0.34 | 5.25 × 10$^{-15}$ | 1 | 16 |
| Co | -0.252 | 3.60 × 10$^{-6}$ | -5.44 | 5.31 | 4.70 × 10$^{-15}$ | 9 | 17 |
| Ni | -0.280 | 2.60 × 10$^{-6}$ | -5.59 | 6.09 | 4.89 × 10$^{-15}$ | 9 | 18 |
| Au (111) | 0.532 | 2.50 × 10$^{-7}$ | -6.60 | 16.05 | 6.86 × 10$^{-15}$ | 1 | 19 |
| Au (111) | 0.532 | 3.38 × 10$^{-7}$ | -6.47 | 16.35 | 6.86 × 10$^{-15}$ | 1 | 20 |
| Poly Au | 0.532 | 1.40 × 10$^{-7}$ | -6.85 | 15.47 | 6.86 × 10$^{-15}$ | 1 | 19 |
| Re | -0.195 | 1.25 × 10$^{-6}$ | -5.90 | 2.30 | 6.04 × 10$^{-15}$ | 9 | 21 |
| Re | -0.195 | 1.00 × 10$^{-6}$ | -6.00 | 2.07 | 6.04 × 10$^{-15}$ | 9 | 22 |
| Bi | 1.099 | 8.00 × 10$^{-10}$ | -9.10 | 33.10 | 1.65 × 10$^{-14}$ | 1 | 23 |
| Cd | 1.031 | 1.7 × 10$^{-11}$ | -10.77 | 25.83 | 7.49 × 10$^{-15}$ | 1 | 24 |
| In | 0.912 | 1.5 × 10$^{-11}$ | -10.82 | 21.33 | 9.30 × 10$^{-15}$ | 1 | 25 |

**Selection of HER Exchange Current Densities from the Research Literature**

Experimental exchange current densities (Table 4) were collected from prior literature reports that showed evidence for a high level of analytical rigor. Each of the following were treated as exclusion criteria by incrementally decreasing a "rigor score" for the associated report:

- Electrolytes were not pre-purified or noted to be of highest available commercial purity
- Counter electrodes comprised materials with higher HER activity than the working electrode; these can dissolve and re-deposit on the working electrode and significantly modify its catalytic activity
- Electrode cleaning protocols (if used) involved exclusions to potentials outside the stability limits for the noted pure metal in strong acid conditions
- Evidence that the working electrode was not completely flat (e.g., roughness factor ≥ 2) and the surface roughness was not taken into consideration in the reported exchange current density
- Tafel plots used for kinetic analysis did not give rise to clearly linear behavior over at least 1 order of magnitude in current density
- Mass transfer limitations convoluted kinetic analysis; note this is especially important for high-performing catalysts like Pt, whose HER activity is so high that conventional hydrodynamic methods like RDE cannot achieve a pure kinetic limit
- Control measurements using comparatively well understood HER catalysts (usually Pt) exhibited excessively low or inconsistent catalytic activity

HER measurements exhibiting one of the deficiencies listed above very often suffered from several, which resulted in a subset of measurements with high rigor and another subset with relatively low rigor. Reports with high rigor are shown in the table, and these were used for the regression analyses in the main text. Notes have also been included in Table 4 summarizing the associated experimental protocols, where the bold text notes relatively minor experimental concerns or incomplete information. Mo and W entries are included in Table 4, but these metals were not included in our analysis because neither is thermodynamically stable as a zerovalent metal under HER conditions in acid; accordingly, DFT-calculated H-binding energies are not directly comparable to experimental results, which most likely involve HER on partially oxidized Mo and W sites.

Several other metals (e.g., Ni and Co) are also only stable in an oxidized form at applied potentials near 0 V vs RHE in strong acid solution, but the oxidation products are soluble (and therefore do not irreversibly modify the electrode surface) and rigorous measurements can be executed over sufficiently negative applied potentials to maintain a metallic composition.

**Table S5.** The collected exchange current densities from experiments with comments.

| Electrode | Reported $j_0$ (A/cm$^2$) | Electrolyte | Temperature | Reference |
|---|---|---|---|---|
| Pt (111) | 4.5×10$^{-4}$ | 0.05 M $H_2SO_4$ | 303 K | 10 |
| Pt (100) | 6.0×10$^{-4}$ | | | |
| Pt (110) | 9.8×10$^{-4}$ | | | |

- Studied different crystal facets of Pt at different temperatures
- Single crystal electrodes tested, RDE measurement
- Low electrolyte concentration chosen to be able to clearly distinguish hydrogen activity
- Electrode surface protected by a drop of water, luggin capillary for reference electrode to avoid Cl$^-$ contamination
- HUPD characterization correlated to the theoretical charge to determine adsorption layers
- Tafel plots determined from the kinetically limited region
- Exchange current densities obtained from micropolarization region
- **$j_0$ might still contain contribution from diffusion**

| Electrode | Reported $j_0$ (A/cm$^2$) | Electrolyte | Temperature | Reference |
|---|---|---|---|---|
| Pt/C | 1.6×10$^{-2}$ | 0.2 M $H_3PO_4$ | 293 K | 11 |

- Studied precious metal catalysts at different pHs
- Commercial powders tested, RDE measurement
- Luggin capillary, Pt counter
- **Performed CVs in 0.1 M KOH prior to testing at different electrolytes (contamination risk)**
- ECSA determined from HUPD peaks in 0.1 M KOH
- **ECSA obtained from HUPD is 1.6 times lower than that obtained by CO-stripping**
- Exchange current densities obtained from Butler Volmer, are consistent with the measurements from $H_2$-pump; **may still be transport limited**

| Electrode | Reported $j_0$ (A/cm$^2$) | Electrolyte | Temperature | Reference |
|---|---|---|---|---|
| Pt/C | 1.2×10$^{-1}$ | 0.1 M $HClO_4$ | 313 K | 12 |

- Studied precious metal catalysts at different temperatures
- Commercial powders were tested in $H_2$ pump configuration (speeds up mass transfer)
- Pt/C counter/reference electrode, scrupulous cell cleaning
- ECSA calculated by CO-stripping at 293 K are consistent with TEM analysis
- $j_0$ values were calculated by Butler Volmer and micro-polarization region, and was within 10 % error

| Electrode | Reported $j_0$ (A/cm$^2$) | Electrolyte | Temperature | Reference |
|---|---|---|---|---|
| Ir/C | 3.6×10$^{-2}$ | 0.1 M $HClO_4$ | 313 K | 12 |

- Studied precious metal catalysts at different temperatures
- Commercial powders were tested in $H_2$ pump configuration
- **Pt/C counter/reference electrode**, rigorous cell cleaning

- ECSA calculated by CO-stripping at 293 K are consistent with TEM analysis
- $j_0$ values were calculated by Butler Volmer and the micro-polarization region and were nearly same. Oxide covering did not have a huge influence.

| Ir/C | $1.28 \times 10^{-2}$ | 0.2 M $H_2SO_4$ | 293 K | 11 |
|---|---|---|---|---|

- Commercial powders tested, RDE measurement
- Luggin capillary, **Pt counter**
- **Performed CVs in 0.1 M KOH prior to testing at different electrolytes**
- ECSA determined from HUPD peaks in 0.1 M KOH
- ECSA obtained from HUPD is almost same as that obtained by CO-stripping
- Exchange current density values obtained from Butler Volmer, are consistent with the measurements from $H_2$-pump; **may still be transport limited**

| Pd | $1.9 \times 10^{-4}$ | 0.5 M $H_2SO_4$ | Not mentioned | 13 |
|---|---|---|---|---|

- Studied nanoporous Pd (powder)
- Carbon counter, nitrogen purge, RDE measurement
- **Current normalization not mentioned**, appears to be from electrode area
- ECSA calculated from $C_{dl}$
- Linear Tafel fitting

| Pd/C | $3.0 \times 10^{-3}$ | 0.1 M $HClO_4$ | 313 K | 12 |
|---|---|---|---|---|

- Commercial powders were tested in $H_2$ pump configuration
- **Pt/C counter/reference electrode**, rigorous cell cleaning
- ECSA calculated by CO-stripping at 293 K are consistent with TEM analysis
- $j_0$ values were calculated by Butler Volmer and the micropolarization region and were nearly same. Hydride covering did not have a significant influence.

| Pd/C | $8.4 \times 10^{-4}$ | 0.1 M $HClO_4$ | 293 K | 11 |
|---|---|---|---|---|

- Commercial powders tested, RDE measurement
- Luggin capillary, **Pt counter**
- **Performed CVs in 0.1 M KOH prior to testing at different electrolytes**
- ECSA determined from PdO peaks in 0.1 M KOH
- **ECSA obtained from HUPD is slightly lower (by 1.2 times) than that obtained by CO-stripping**
- $j_0$ values obtained from Butler Volmer, consistent with the measurements from $H_2$-pump

| Rh/C | $5.2 \times 10^{-3}$ | 0.1 M $HClO_4$ | 313 K | 12 |
|---|---|---|---|---|

- Commercial powders were tested in $H_2$ pump configuration
- **Pt/C counter/reference electrode**, rigorous cell cleaning
- ECSA calculated by CO-stripping at 293 K are consistent with TEM analysis
- $j_0$ values were calculated by Butler Volmer and the micropolarization region and were nearly same. Oxide covering did not have a significant influence.

| Rh/C | $6.7 \times 10^{-3}$ | 0.1 M $HClO_4$ | 293 K | 11 |
|---|---|---|---|---|

- Commercial powder tested, RDE measurement
- Luggin capillary, **Pt counter**
- **Performed CVs in 0.1 M KOH prior to testing at different electrolytes**

- Currents normalized by ECSA measured in 0.1 M KOH
- **ECSA obtained from HUPD is 1.8 times lower than that obtained by CO-stripping**
- Exchange current density values obtained from Butler Volmer, are consistent with the measurements from $H_2$-pump

| Ru | $4.5 \times 10^{-3}$ | 1M HCl + NaCl | 298 K | 15 |
|---|---|---|---|---|

- Studied Ru cylinder, mounted on ptfe cup
- **Pt counter**, separated from working electrode using frit
- Pre-electrolysis performed but **conditions are unclear** (pre-electrolysis implies rigorous purification)
- After pre-electrolysis potential sequence of +1 V vs RHE for 10 s followed by -1 V for 10 mins was repeated 6 times with final cathodic pulse for 10 minutes. **Rest potential was observed to be -1 V vs RHE.**

| Cu | $1.45 \times 10^{-7}$ | 0.1 N HCl | Not mentioned | 16 |
|---|---|---|---|---|

- Wire working electrode
- Graphite counter
- Pre-electrolysis was performed for several hours; **HCl electrolyte may allow for some Cu dissolution**
- Statistical analysis included

| Co | $3.6 \times 10^{-6}$ | 1 M $H_2SO_4$ | 293 K | 26 |
|---|---|---|---|---|

- Studied rod electrode, electrolytically polished in $H_3PO_4$
- Detailed cleaning procedure followed
- Cathodically pre-polarized starting from low current density
- Overpotential increased by applying cathodic current or with several hours of electrolyte contact
- Also calculated Tafel slope for the dissolution process, consistent with prior literature

| Ni | $2.6 \times 10^{-6}$ | 0.5 M $H_2SO_4$ | 295 K | 18 |
|---|---|---|---|---|

- Studied electrodeposited Ni as control, **Ni could be coated with $Ni^{2+}$ compounds**
- **Pt counter** separated from main cell using glass frit, argon purge
- Electrode was polarized at HER potentials to remove surface oxides
- Tafel plot measured in kinetically controlled regime
- **Tafel slope is higher than theoretical value; attributed to Ni oxidation**
- $j_0$ obtained from tafel plot

| Au (111) | $2.5 \times 10^{-7}$ | | | |
|---|---|---|---|---|
| Au (100) | $0.5 \times 10^{-7}$ | 0.1 M $HClO_4$ | Not mentioned | 19 |
| Au (110) | $0.3 \times 10^{-7}$ | | | |

- Studied single crystals with different crystal facets
- Hanging meniscus rotating disc technique, nitrogen purge, Au counter electrode
- Surface of electrode protected with electrolyte drop
- HER activity was independent of the potential history (scanning even to oxidation potentials), contrary to literature, owing to cleaner surfaces and solutions
- **Did not document detailed cell cleaning protocols**

| | | | | |
|---|---|---|---|---|
| | | | | |

- **Tafel slope in the low potential region is reported (< 150 mV)**

| Au (111) | 3.38×10$^{-7}$ | 0.5 M H$_2$SO$_4$ | Not mentioned | 20 |
|---|---|---|---|---|

- Studied single crystal as control
- **Pt counter**, nitrogen purge, hanging meniscus
- **Current normalized to geometric area; no ECSA**

| Poly Au | 1.40×10$^{-7}$ | 0.1 M HClO$_4$ | Not mentioned | 19 |
|---|---|---|---|---|

- Hanging meniscus rotating disc technique, nitrogen purge, Au counter electrode
- Surface of electrode protected with electrolyte drop
- HER activity was independent of the potential history (scanning even to oxidation potentials) owing to cleaner surfaces and solutions
- **Did not document detailed cell cleaning protocols**
- **Tafel slope in the low potential region is reported (< 150 mV)**

| Re | 1.25×10$^{-6}$ | 0.5 M H$_2$SO$_4$ | 298 K | 21 |
|---|---|---|---|---|

- Polished wire working electrode
- **Pt counter**, hydrogen purge
- Native surface oxide formation was minimized by polarizing at -0.4 V vs NHE
- **Tafel fit included narrow range at low overpotential (-0.11 to -0.2 V vs NHE)**

| Re | 1×10$^{-6}$ | 0.5 M H$_2$SO$_4$ | 298 K | 22 |
|---|---|---|---|---|

- Polished wire working electrode
- **Pt counter**, hydrogen purge
- Polarized at -0.1 V vs RHE, claim to have metallic Re
- **Current normalized to geometric area; no ECSA**
- 

| Cd | 1.7×10$^{-11}$ | 0.5 N H$_2$SO$_4$ | Not mentioned | 24 |
|---|---|---|---|---|

- Metal wire working electrode
- Heated electrode in hydrogen
- Detailed cleaning procedure followed
- Electrolyte was purged with pre-purified nitrogen to remove excess oxygen and then purged with hydrogen
- Electrolyte was further purified by pre-electrolysis at 1 mA/cm$^2$ for 15 – 20 hours

| Bi | 8×10$^{-10}$ | 4.8 M H$_2$SO$_4$ | Not mentioned | 23 |
|---|---|---|---|---|

- Polished metal wire working electrode
- **Pt counter electrode**
- Held the potential at HER potential for 10 mins prior to analysis
- Tafel plot measured in kinetically controlled regime

| In | 1.51×10$^{-11}$ | 0.1 M HClO$_4$ | 303 K | 25 |
|---|---|---|---|---|

- Cylindrical working electrode
- Electropolished at negative potential to remove oxide layer before analysis
- **Pt counter and reference electrode**, Luggin capillary used

- Varied electrode treatment conditions and electrolyte concentration

Table S6. Ten best models identified by SISSO. Primary features used are atomic radius ($R$), atomic number (N), atomic mass (M) period in Periodic Table (P), metal density ($\rho$), work function ($\phi$), electron affinity ($E_A$), ionization energy (*I*), Pauling electronegativity ($\chi$), and hydrogen adsorption energy ($\Delta G_H$).

| SISSO Model | $r^2$ |
|---|---|
| $((\chi+(\Delta G_H+\chi))+((P\Delta G_H)(\Delta G_H/\chi)))$ | 0.9795 |
| $(((N\chi)(\chi/M))+((\Delta G_H E_A)(\Delta G_H/\chi)))$ | 0.9789 |
| $(((PR)(P/M))((PR)(\Delta G_H+\chi)))$ | 0.9782 |
| $(((M+N)(\chi/M))+((P\Delta G_H)(\Delta G_H/\chi)))$ | 0.9775 |
| $((\chi(R\chi))+((R\Delta G_H)(\Delta G_H+E_A)))$ | 0.9768 |
| $(((\Delta G_H+\chi)(\Delta G_H/\chi))((\Delta G_H/\chi)+(\chi/\Delta G_H)))$ | 0.9766 |
| $((R(\Delta G_H+\chi))+((R\Delta G_H)(\Delta G_H/\chi)))$ | 0.9762 |
| $(((\Delta G_H+\phi)(\chi/\phi))/((\chi/\phi)+(\phi/\chi)))$ | 0.9759 |
| $(((R\Delta G_H)(\Delta G_H+E_A))+((R\chi)(\Delta G_H+\chi)))$ | 0.9753 |
| $((\Delta G_H(\Delta G_H+\chi))((EA/\chi)+(\chi/\Delta G_H)))$ | 0.9751 |